\pdfoutput=1
\documentclass[prl,reprint,superscriptaddress,showpacs,nobalancelastpage]{revtex4-1}
\usepackage{amssymb,amsmath}
\usepackage{graphicx}
\usepackage[small]{subfigure}
\usepackage{mathptmx}
\usepackage{fancyhdr}
\usepackage{CJK}
\usepackage[breaklinks=true,colorlinks=true,urlcolor=blue,
citecolor=blue,linkcolor=blue,bookmarks=false]{hyperref}

\def\unit#1{\mathord{\thinspace\rm #1}}
\def\func#1{\mathop{\rm #1}\nolimits}%
\subfiguretopcaptrue

\fancypagestyle{plain}{
\fancyhf{}
\fancyfoot[C]{\small \thepage}
\fancyhead[C]{\small Published version: \href{http://journals.aps.org/prl/abstract/10.1103/PhysRevLett.114.036601}{Phys.\ Rev.\ Lett.\ \textbf{114}, {036601} (2015)}}
}

\begin{document}

\begin{CJK*}{Bg5}{bsmi}

\title{Scalable Tight-Binding Model for Graphene}

\author{Ming-Hao Liu (¼B©ú»¨)}

\email{minghao.liu.taiwan@gmail.com}

\affiliation{Institut f\"ur Theoretische Physik, Universit\"at Regensburg, D-93040 Regensburg, Germany}

\author{Peter Rickhaus}

\affiliation{Department of Physics, University of Basel, Klingelbergstrasse 82, CH-4056 Basel, Switzerland}

\author{P\'eter Makk}

\affiliation{Department of Physics, University of Basel, Klingelbergstrasse 82, CH-4056 Basel, Switzerland}

\author{Endre T\'ov\'ari}

\affiliation{Department of Physics, Budapest University of Technology and Economics and Condensed Matter Research Group of the Hungarian Academy of Sciences, Budafoki ut 8, 1111 Budapest, Hungary}

\author{Romain Maurand}

\affiliation{University Grenoble Alpes and CEA-INAC-SPSMS, F-38000 Grenoble, France}

\author{Fedor Tkatschenko}

\affiliation{Institut f\"ur Theoretische Physik, Universit\"at Regensburg, D-93040 Regensburg, Germany}

\author{Markus Weiss}

\affiliation{Department of Physics, University of Basel, Klingelbergstrasse 82, CH-4056 Basel, Switzerland}

\author{Christian Sch\"onenberger}

\affiliation{Department of Physics, University of Basel, Klingelbergstrasse 82, CH-4056 Basel, Switzerland}

\author{Klaus Richter}

\affiliation{Institut f\"ur Theoretische Physik, Universit\"at Regensburg, D-93040 Regensburg, Germany}

\date{\today}

\begin{abstract}

Artificial graphene consisting of honeycomb lattices other than the atomic layer of carbon has been shown to exhibit electronic properties similar to real graphene. Here, we reverse the argument to show that transport properties of real graphene can be captured by simulations using \textquotedblleft theoretical artificial graphene\textquotedblright. To prove this, we first derive a simple condition, along with its restrictions, to achieve band structure invariance for a scalable graphene lattice. We then present transport measurements for an ultraclean suspended single-layer graphene {\it pn} junction device, where ballistic transport features from complex Fabry-P\'erot interference (at zero magnetic field) to quantum Hall effect (at unusually low field) are observed, and are well reproduced by transport simulations based on properly scaled single-particle tight-binding models. Our findings indicate that transport simulations for graphene can be efficiently performed with a strongly reduced number of atomic sites, allowing for reliable predictions for electric properties of complex graphene devices. We demonstrate the capability of the model by applying it to predict so-far unexplored gate-defined conductance quantization in single-layer graphene.

\end{abstract}


\maketitle

\end{CJK*}

\thispagestyle{plain}

Graphene is a promising material for its special electrical, optical, thermal, and mechanical properties. In particular, the conic electronic structure that mimics two-dimensional (2D) massless Dirac fermions has attracted much attention on both the academic and industrial side. Soon after the \textquotedblleft debut\textquotedblright\ of single-layer graphene \cite{Novoselov2004,Berger2004} and the subsequent confirmation of its relativistic nature \cite{Zhang2005,Novoselov2005,Li2007}, the exploration of Dirac fermions in condensed matter has been further extended to honeycomb lattices other than graphene, including optical lattices \cite{Zhu2007,Wunsch2008,Uehlinger2013}, semiconductor nanopatterning \cite{Park2009,Gibertini2009,Rasanen2012,Kalesaki2014}, molecular arrays on Cu(111) surfaces \cite{Gomes2012}, or even macroscopic, dielectric resonators for microwave propagation \cite{Kuhl2010,Bellec2013}, all of which have been shown to exhibit similar electronic properties as real graphene and hence are referred to as \emph{artificial graphene} \cite{Polini2013}.

\begin{figure}[b]
\includegraphics[width=0.9\columnwidth]{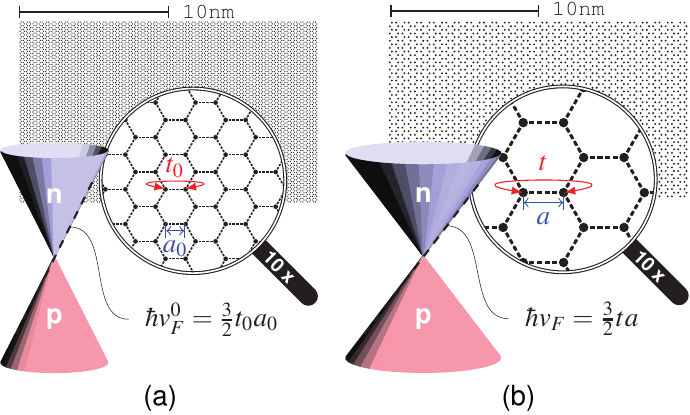}
\caption{(Color online) Schematic of a sheet of (a) real graphene and (b) scaled graphene and their conical low-energy band structures. In (a), the lattice spacing $a_{0}\approx 0.142\unit{nm}$, the hopping energy $t_{0}\approx 2.8\unit{eV}$, and the Fermi velocity $v_{F}^{0}\approx 10^{8}\unit{cm}\unit{s^{-1}}$.} 
\label{fig1}
\end{figure}

Here, we reverse the argument to show that transport properties of real graphene can be captured by simulations using \textquotedblleft theoretical artificial graphene\textquotedblright, by which we mean a honeycomb lattice with its lattice spacing $a$ different from the carbon-carbon bond length $a_{0}$ of real graphene; see Fig.\ \ref{fig1}. From a theoretical point of view, this can be achieved only if the considered theoretical artificial lattice, which will be shortly referred to as artificial graphene or scaled graphene, has its energy band structure identical to that of real graphene.
In this paper, we first derive a simple condition, along with its restrictions, to achieve the band structure invariance of graphene with its bond length scaled from $a_{0}$ to $a$, even in the presence of magnetic field. We then prove the argument by presenting transport measurements for an ultraclean suspended single-layer graphene {\it pn} junction device, where ballistic transport features from Fabry-P\'erot interference to quantum Hall effect are observed, and are well reproduced by quantum transport simulations based on the scaled graphene. To go one step further, we demonstrate the capability of the scaling approach by applying it to uncover one of the experimentally feasible yet unexplored transport regimes: gate-defined zero-field conductance quantization.

We begin our discussion with the standard tight-binding model for 2D graphite \cite{Wallace1947}, i.e., bulk graphene, and focus on the low-energy range ($|E|\lesssim 1\unit{eV}$) which is addressed in most graphene transport measurements. In this regime, the effective Dirac Hamiltonian $H_{\text{eff}}=v_{F}\vec{\sigma}\cdot \mathbf{p}$ associated with the celebrated linear band structure $E(\mathbf{k})=\pm \hbar v_{F}|\mathbf{k}|$ describes the graphene system well. Here $v_{F}\approx 10^{8}\unit{cm}\unit{s}^{-1}$ is the Fermi velocity in graphene, and $\hbar \mathbf{k}$, the eigenvalue of the operator $\vec{\sigma}\cdot \mathbf{p}$ [Pauli matrices $\vec{\sigma}=(\sigma_x,\sigma_y)$ act on the pseudospin properties], is the quasimomentum with $\mathbf{k}$ defined relative to the $K$ or $K^{\prime }$ point in the first Brillouin zone. In terms of the tight-binding parameters, one replaces $\hbar v_{F}$ with $(3/2)t_{0}a_{0}$, where $t_{0}\approx 2.8\unit{eV}$ is the nearest neighbor hopping parameter and $a_{0}\approx 0.142\unit{nm}$ is the lattice site spacing, i.e., $E_0(k)=(3/2)t_{0}a_{0}k$ for real graphene \footnote{Note that the next nearest neighbor hopping $t^{\prime }$ does not play a role for describing the low-energy physics of graphene, and will not be considered in this work.}. Now, we consider the scaled graphene described by the same tight-binding model but with hopping parameter $t$ and lattice spacing $a$, and introduce a scaling factor $s_{f}$ such that $a=s_{f}a_{0}$. The real and scaled graphene sheets along with their low-energy band structures are schematically sketched in Fig.\ \ref{fig1}. The low-energy dispersion for scaled graphene is naturally expected to be $E(k)=(3/2)tak$. Thus to keep the energy band structure unchanged while scaling up the bond length by a factor of $s_{f}$, the condition
\begin{equation}
a=s_{f}a_{0},\qquad t=\frac{t_{0}}{s_{f}}.  \label{scaling condition}
\end{equation}
becomes self-evident.

Clearly, Eq.\ \eqref{scaling condition} applies only when the linear approximation is valid. In terms of the long wavelength limit, this means that the Fermi wavelength should be much longer than the lattice spacing: $\lambda _{F}\gg a$, from which using Eq.\ \eqref{scaling condition} the following validity criterion can be deduced:
\begin{equation}
s_{f}\ll \frac{3t_{0}\pi }{\left\vert E_{\max }\right\vert },
\label{scaling criterion}
\end{equation}%
where $E_{\max }$ is the maximal energy of interest for investigating a particular real graphene system. Considering graphene on typical Si/300nm SiO$_2$ substrate, the usually accessed carrier density range is less than $10^{13}\unit{cm}^{-2}$ \cite{Novoselov2004}. This implies that the energy range of interest lies within $|E_{\max }|\lesssim 0.4\unit{eV}$, leading to $s_{f}\ll 66$ from Eq.\ \eqref{scaling criterion} (using $t_{0}=2.8\unit{eV}$). For suspended graphene, typical carrier densities can hardly reach $10^{12}\unit{cm}^{-2}$ \cite{Bolotin2008}, so that $|E_{\max }|\lesssim 0.1\unit{eV}$ allows for a larger range of the scaling factor, $s_{f}\ll 264$.

In the presence of an external magnetic field, the Peierls substitution \cite{Peierls1933} is the standard method to take into account the effect of a uniform out-of-plane magnetic field $B_{z}$ within the tight-binding formulation. In addition to the long wavelength limit \eqref{scaling criterion}, the validity of the Peierls substitution, however, imposes a further restriction for the scaling \cite{Goerbig2011}: $l_{B}\gg a$, where $l_{B}=\sqrt{\hbar /eB_{z}}$ is the magnetic length. In terms of $a=s_{f}a_{0}$ given in Eq.\ \eqref{scaling condition}, this restriction reads%
\begin{equation}
s_{f}\ll \frac{l_{B}}{a_{0}}\approx \frac{180}{\sqrt{B_{z}}},
\label{scaling restriction}
\end{equation}%
where $B_{z}$ is in units of $\unit{T}$. Equations \eqref{scaling condition}--\eqref{scaling restriction} complete the description of band structure invariance for scaled graphene.

\begin{figure}[tbp]
\includegraphics[width=0.95\columnwidth]{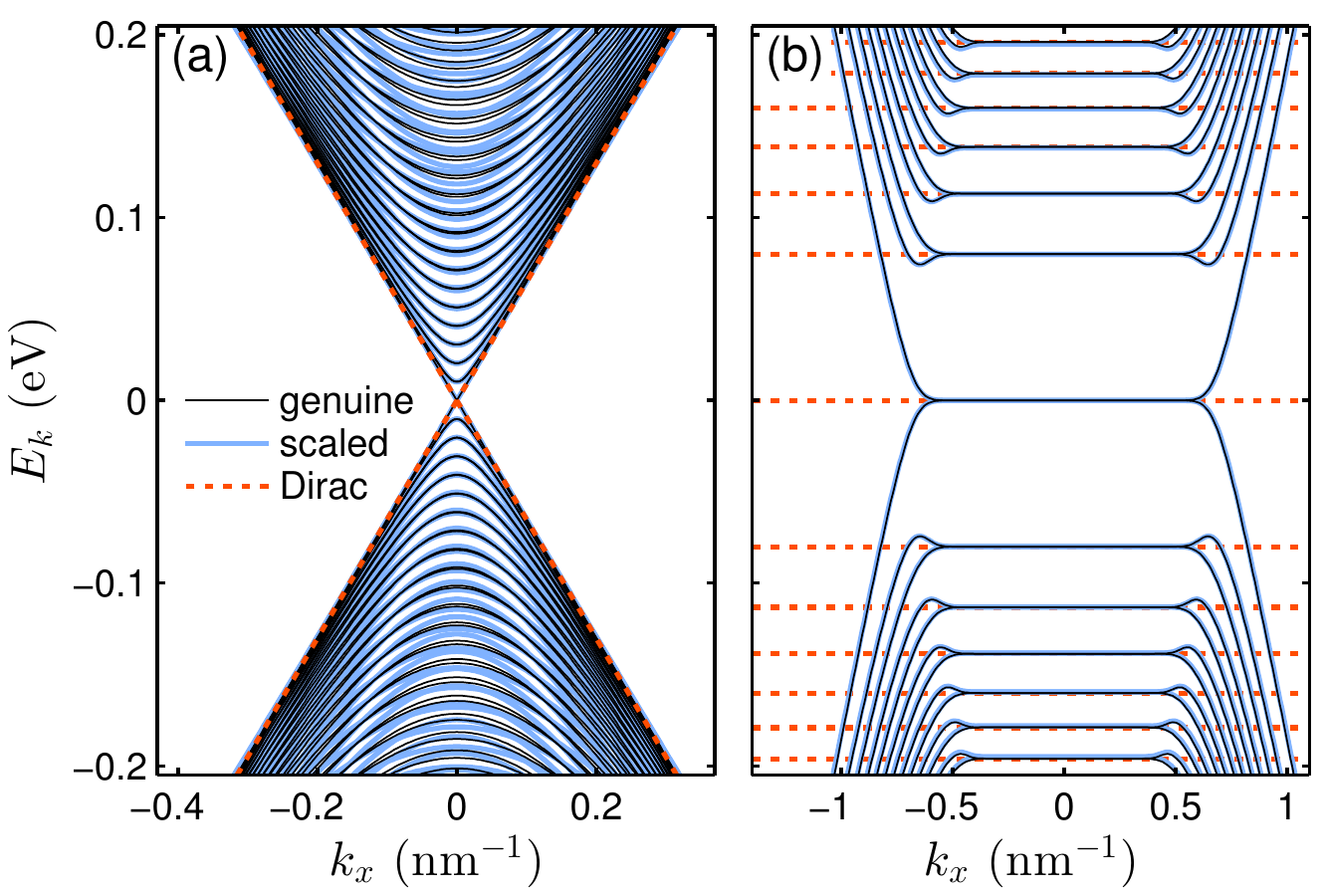}
\caption{(Color online) Band structure consistency check using an armchair graphene ribbon with width $200\unit{nm}$, (a) in the absence of magnetic field, and (b) in the presence of a uniform magnetic field $B_z=5\unit{T}$. The comparison is done for both (a) and (b) between the genuine graphene with $s_f=1$ and scaled graphene with $s_f=4$, which correspond to chain numbers $N_a=800$ and $N_a=200$, respectively.}
\label{fig2}
\end{figure}

The above discussion is based on bulk graphene, but the listed conditions apply equally well to finite-width graphene ribbons. To show a concrete example of band structure invariance under scaling, we consider a 200-nm-wide armchair ribbon and compare the band structures of the genuine case with $s_{f}=1$ and the scaled case with $s_{f}=4$ in Fig.\ \ref{fig2}(a) for $B_z=0$. The scaled graphene band structure well matches the genuine one at low energy $|E|\lesssim 0.1\unit{eV}$, and starts to deviate at higher energy but stays rather consistent within the shown energy range of $|E|\leq 0.2\unit{eV}$. Both of the band structures are well bound by the linear Dirac model that corresponds to the bulk graphene. The band structure invariance remains true when a magnetic field is applied, as seen in Fig.\ \ref{fig2}(b), where $B_{z}=5\unit{T}$ is considered. The pronounced flat bands in both cases match perfectly with the relativistic Landau levels $E_{n_{L}}=\func{sgn}(n_{L})\sqrt{2eB_{z}\hbar v_{F}^{2}|n_{L}|}$ solved from the Dirac model \cite{Zhang2005,Novoselov2005,Li2007,Goerbig2011}, where $n_L=0,\pm 1,\pm 2, \cdots$. The band structure invariance based on Eqs.\ \eqref{scaling condition}--\eqref{scaling restriction} can be easily shown to hold also for zigzag graphene ribbons.

\begin{figure}[t]
\includegraphics[width=\columnwidth]{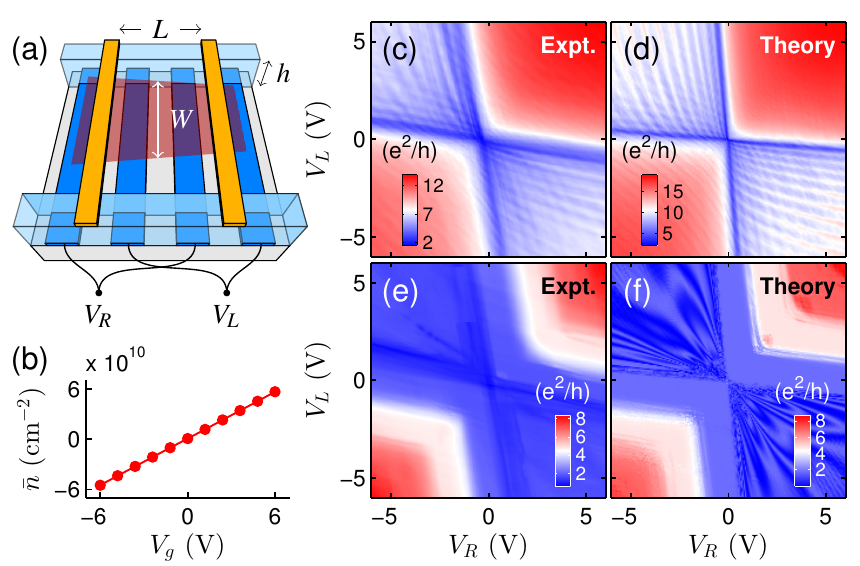}
\caption{(Color online) (a) Sketch of the suspended graphene \textit{pn} junction device with suspension height $h=600\unit{nm}$, contact spacing $L=1680\unit{nm}$, and average flake width $W=2125\unit{nm}$. (b) Mean carrier density as a function of $V_g=V_L=V_R$, based on a 3D electrostatic simulation. Experimental/theoretical data of the two-terminal conductance at (c)/(d) $B_z=0$ and (e)/(f) $B_z=0.2\unit{T}$. Both of (c) and (e) were measured at temperature $T=1.4\unit{K}$, while the simulations were done at zero temperature using (d) $s_f=100$ and (f) $s_f=50$ scaled graphene.}
\label{fig3}
\end{figure}

Having demonstrated that under proper conditions \eqref{scaling condition}--\eqref{scaling restriction} the scaled graphene band structure can be identical to that of real graphene, we next perform quantum transport simulations for a real graphene device, using the scaled graphene. To this end, we have fabricated ultraclean suspended graphene \textit{pn} junctions as sketched in Fig.\ \ref{fig3}(a). First, bottom gates were prepatterned on a Si wafer with $300\unit{nm}$ SiO$_2$ oxide. Afterwards, the wafer was spin-coated with lift-off resist (LOR), and the graphene was transferred on top following the method described in Ref.\ \onlinecite{Dean2010}. Palladium contacts to graphene were made by e-beam lithography and thermal evaporation, and the device was suspended by exposing and developing the LOR resist. Finally, the graphene was cleaned by current annealing at $4\unit{K}$. The fabrication method is described in Refs.\ \cite{Tombros2011,Maurand2014} in detail.

Following the device design of our experiment [sketched in Fig.\ \ref{fig3}(a)], we first build a three-dimensional (3D) electrostatic model to obtain the self-partial capacitances \cite{Cheng1989,Liu2013} of the individual metal contacts and bottom gates, which are computed by the finite-element simulator {\sc FEniCS} \cite{FEniCS} combined with the mesh generator {\sc Gmsh} \cite{gmsh}. The extracted self-partial capacitances from the electrostatic simulation provide us the realistic carrier density profile \footnote{See Supplemental Material for numerical examples of the carrier density profile $n(x,y)$ simulated for the device, the carrier density as a function of energy and magnetic field $n(E,B_{z})$ using scaled graphene ribbons, unipolar quantum Hall data for the measurement and simulation, evaluation of the gate efficiency from the Landau fan diagram, and comments on the speed-up and bilayer graphene.} $n(x,y)$ at any combination of the left and right bottom gate voltages, $V_L$ and $V_R$, respectively. In Fig.\ \ref{fig3}(b), we plot the mean carrier density $\bar{n}$ averaged over the whole suspended graphene region as a function of $V_g=V_L=V_R$. The slope reveals a charging efficiency of the connected bottom gates of about $10^{10}\unit{cm}^{-2}\unit{V}^{-1}$, which is slightly lower than the experimental value of $1.24\times 10^{10}\unit{cm}^{-2}\unit{V}^{-1}$ extracted from the unipolar quantum Hall data \cite{Note2}.

In the absence of magnetic field, the Fermi energy as a function of carrier density within the low-energy range can be well described by the Dirac model, $E(n)=\func{sgn}(n)\hbar v_{F}\sqrt{\pi |n|}$. This suggests: for a given carrier density at $(x,y)$, applying a local energy band offset defined by $V(x,y)=-\func{sgn}[n(x,y)]\hbar v_{F}\sqrt{\pi |n(x,y)|}$ guarantees that the locally filled highest level fulfills the amount of the simulated carrier density $n(x,y)$ and is globally fixed at $E=0$ for all $(x,y)$. We therefore consider the model Hamiltonian,%
\begin{equation}
H_{\text{model}}=\sum_{i}V(x_{i},y_{i})c_{i}^{\dag}c_{i}-t(s_{f})\sum_{\langle i,j\rangle }c_{i}^{\dag }c_{j},  \label{Hmodel}
\end{equation}%
and apply the Landauer-B\"{u}ttiker formalism \cite{Datta1995} to calculate the transmission function $T$ at energy $E=0$ and temperature zero. In Eq.\ \eqref{Hmodel}, the indices $i$ and $j$ run over the lattice sites within the scattering region defined by an artificial graphene scaled by $s_{f}$, and the second term contains the nearest neighbor hopping elements with strength $t(s_{f})$ given in Eq.\ \eqref{scaling condition}.

For zero-field transport, we compute the conductance map $G(V_R,V_L)$ from the transmission function $T$ using $G=(e^{2}/h)[T^{-1}+R_{c}/(h/e^{2})]^{-1}$, where the contact resistance is deduced from the quantum Hall measurement to be $R_{c}\approx 1080\unit{\Omega}\approx 4.2\times 10^{-2}(h/e^{2})$. The measured and simulated conductance maps are reported in Figs.\ \ref{fig3}(c) and \ref{fig3}(d), respectively, both exhibiting two overlapping sets of Fabry-P\'{e}rot interference patterns in the bipolar blocks similar to Refs.\ \cite{Grushina2013,Rickhaus2013}. Strikingly, the theory data reported in Fig.\ \ref{fig3}(d) is based on a scaled graphene with $s_{f}=100$ because of the rather low density (energy) in our ultraclean device. From the estimated maximal carrier density [Fig.\ \ref{fig3}(b)], we find $|E_{\max }|\approx 28\unit{meV}$, such that Eq.\ \eqref{scaling criterion} roughly gives $s_{f}\ll 10^{3}$, suggesting that $s_{f}=100$ is acceptable. Simulations with smaller $s_{f}$ have been performed and do not significantly differ from the reported map. 

To correctly account for the magnetic field effect in the transport simulation, the first step, similar to the zero-field case, is to extract the proper energy band offset from the given carrier density through the carrier-energy relation, for which an exact analytical formula does not exist. Numerically, the carrier density as a function of energy and magnetic field, $n(E,B_{z})$, can be computed also using the Green's function method \cite{Note2}, and subsequently provide $E(n,B_{z})$. The desired energy band offset is then again given by the negative of it. Thus the magnetic field in the transport simulation requires, in addition to the Peierls substitution of the hopping parameter, the modification on the on-site energy term of Eq.\ \eqref{Hmodel}, $V(x_{i},y_{i})\rightarrow V(x_{i},y_{i};B_{z})=-E(n(x_{i},y_{i}),B_{z})$, where $n(x_{i},y_{i})$ is obtained from the same electrostatic simulation and is assumed to be unaffected by the magnetic field, i.e., we assume the electrostatic charging ability of the bottom gates is not influenced by the magnetic field.

At field strength $B_{z}=0.2\unit{T}$, Fig.\ \ref{fig3}(e) shows the measured conductance map and is qualitatively reproduced by the simulation Fig.\ \ref{fig3}(f) done by an $s_f=50$ scaled graphene in the presence of weak disorder. We observe very good agreement in the conductance range as well as in the conductance features in the unipolar blocks. In the bipolar blocks, however, the simulation reveals a fine structure that is found to be sensitive to spatial and edge disorder, but is not observed in the present experimental data. Nevertheless, the conductance in the bipolar blocks varies between $0$ and $2e^{2}/h$ in both experiment and theory, and neither of them exhibits the fractional plateaus \cite{Abanin2007}. Thus the bipolar blocks of Figs.\ \ref{fig3}(e) and \ref{fig3}(f) reveal a conductance behavior due to the ballistic smooth graphene \textit{pn} junctions very different from the diffusive sharp ones \cite{Williams2007,Ozyilmaz2007,Long2008}. Note that here we have considered Anderson-type disorder by adding to the model Hamiltonian \eqref{Hmodel} the potential term $\sum_{i}U_{i}c_{i}^{\dag}c_{i}$, where $U_i$ is a random number $U_{i}\in [-U_{\rm dis}/2,U_{\rm dis}/2]$ with disorder strength $U_{\rm dis}=6\unit{meV}$ used in the theory map of Fig.\ \ref{fig3}(f). The quantized conductance in the unipolar blocks of the simulated map is found to be robust against the disorder potential, whose quantitative effect is yet to be established for the scaled graphene and is beyond the scope of the present discussion.

\begin{figure}[t]
\includegraphics[width=\columnwidth]{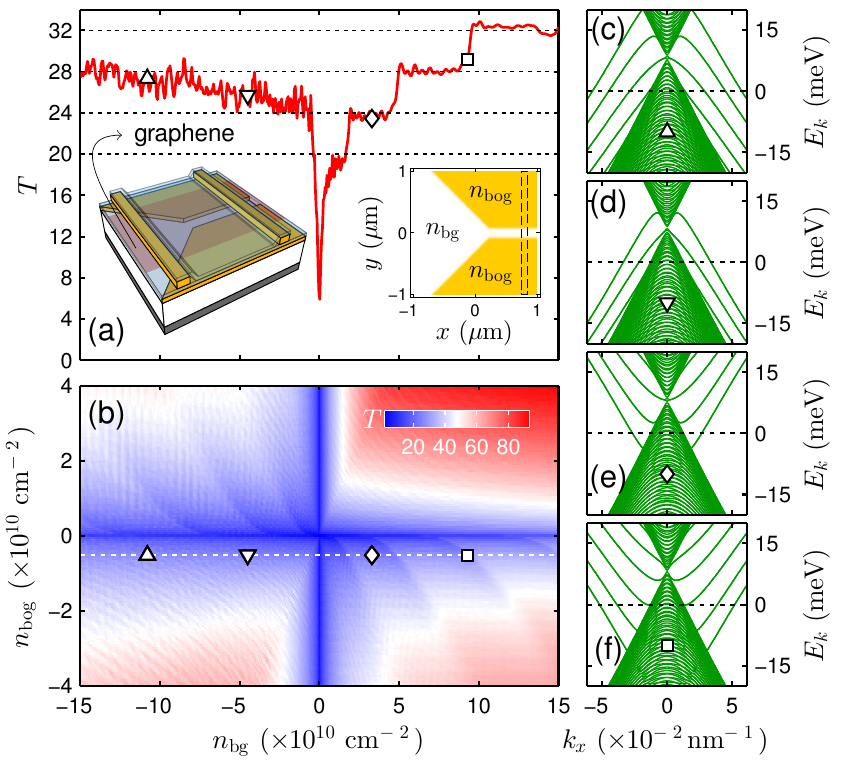}
\caption{(Color online) (a) Two-terminal transmission $T$ as a function of backgate carrier density $n_{\rm bg}$ at fixed bottom gate carrier density $n_{\rm bog}$ for a hBN-sandwitched ballistic graphene device (left inset), assuming a flake size of $2\times 2\unit{\mu m}^2$ subject to the carrier density profile sketched in the right inset. (b) 2D map of the transmission function $T(n_{\rm bg},n_{\rm bog})$; the white dashed line indicates the line trace of (a). (c)--(f) Band structures computed by a unit cell cut from the right part of the same model flake, as marked by the dashed stripe shown in the right inset of (a). The carrier density configurations of panels (c)--(f) are indicated by the symbols ($\vartriangle,\triangledown,\Diamond$ and $\square$) marked at middle corresponding to those marked in (a) and (b). Both of $T$ and $E_k$ are computed based on clean armchair graphene scaled by $s_f=50$.}
\label{fig4}
\end{figure}

Finally, we apply the scaling approach to uncover one of the experimentally feasible but unexplored transport regimes: gate-defined zero-field conductance quantization of single-layer graphene. We consider a ballistic graphene device with encapsulation of hexagonal boron nitride (hBN) \cite{Dean2010,Wang2013} subject to a global backgate and a pair of trapezoidal bottom gates, forming a 150-nm-wide gate-defined channel in the right part of the graphene flake. The device layout is sketched in the left inset of Fig.\ \ref{fig4}(a). Due to the screening of the bottom gates, the carrier density in the bottom-gated region, $n_{\rm bog}$, and the backgated region, $n_{\rm bg}$, can be independently controlled. The ideal carrier density profile within the modeled $2\times 2\unit{\mu m}^2$ flake is shown in the right inset of Fig.\ \ref{fig4}(a), where the left and right leads are attached at $x=\pm 1\unit{\mu m}$.

In the unipolar configuration ($n_{\rm bg} n_{\rm bog}>0$), electrons can freely tunnel between the bottom- and back-gated regions, such that no conductance quantization is expected. In the bipolar configuration ($n_{\rm bg} n_{\rm bog}<0$), however, Klein collimation \cite{Cheianov2006} suppresses oblique tunneling across the pn interfaces, separating the conduction through the narrow channel from that through the $n_{\rm bog}$-region, and the total conductance is expected to vary in discrete quanta of $4e^2/h$ (valley and spin degeneracies) when tuning the channel density $n_{\rm bg}$. This is indeed observed in the 2D map of the transmission function $T(n_{\rm bg},n_{\rm bog})$ reported in Fig.\ \ref{fig4}(b), assuming fixed density of $1.5\times 10^{11}\unit{cm}^{-2}$ in the left and right leads (mimicking \textit{n}-doping contacts). A line cut at $n_{\rm bog}\approx -5.14\times 10^9\unit{cm}^{-2}$ is shown in Fig.\ \ref{fig4}(a), where a clear profile of the quantized conductance plateaus in the bipolar regime can be seen. 

Contrary to the reported signatures of quantized conductance of graphene nanoribbons \cite{Lin2008} and suspended graphene nanoconstrictions \cite{Tombros2011a}, the proposed scheme here is based on a flexible and tunable way of electrical gating using unetched wide graphene such that no localization is expected, and the fabrication process does not require any poorly controlled etching or electrical burning process. In addition, the conductance plateaus predicted here have a rather different origin compared to the usual size quantization (e.g., \cite{Peres2006}). This is illustrated by showing the band structure, considering a unit cell cut from the right part of the same model flake [marked by the dashed stripe in the right inset of Fig.\ \ref{fig4}(a)]. Examples of the resulting hybrid band structures are shown in Figs.\ \ref{fig4}(c)--\ref{fig4}(f), each composed of a dense Dirac cone from the outer wide ($n_{\rm bog}$) region and discrete bands from the inner narrow ($n_{\rm bg}$) region. The former is responsible for a background contribution to the total $T$ leading to a conductance minimum well above zero (contrary to, e.g., \cite{Tombros2011a}), and the latter influences $T$ in a different way depending on the relative polarities of the two regions. In the unipolar examples of Figs.\ \ref{fig4}(c) and \ref{fig4}(d), bands of the two regions mix together, such that $T$ changes continuously. In the bipolar examples of Figs.\ \ref{fig4}(e) and \ref{fig4}(f), $T$ changes abruptly whenever a discrete band is newly populated or depopulated [such as Fig.\ \ref{fig4}(f)].

In conclusion, we have shown that the physics of real graphene can be well captured by studying properly scaled, artificial graphene. This important fact indicates that the number of lattice sites required in transport simulations for graphene based on tight-binding models need not be as massive as in actual graphene sheets. The scaling parameter $s_{f}$, also applicable to bilayer graphene \cite{McCann2013}, scales down the amount of the Hamiltonian matrix elements of the simulated graphene flake by a factor of $s_{f}^{-4}$, and hence strongly reduces the computation overhead \cite{Note2}, making previously prohibited micron-scale 2D devices accessible to accurate simulations. Our findings advance the power of quantum transport simulations for graphene in a simpler and more natural way as compared to the finite-difference method for massless Dirac fermions \cite{Tworzydlo2008}, allowing for reliable predictions for electric properties of complex graphene devices. The illustrated example of applying the scaled graphene to explore one of the new transport regimes---gate-defined zero-field conductance quantization---can be one of the next challenges for graphene transport experiments.

\begin{acknowledgements}
We thank J.\ Bundesmann, S.\ Essert, and V.\ Krueckl and J.\ Michl for valuable suggestions. Financial support by the Deutsche Forschungsgemeinschaft within programs GRK 1570 and SFB 689, by the Hans B\"ockler Foundation, by the Swiss National Science Foundation, the EU FP7 project SE2ND, the ERC Advanced Investigator Grant QUEST, the ERC 258789 is acknowledged. The Swiss National Centres of Competence in Research Quantum Science and Technology (NCCR QSIT), and Graphene Flagship are gratefully acknowledged. 
\end{acknowledgements}

\bibliographystyle{apsrev4-1}
\bibliography{mhl2}

\appendix

\renewcommand{\thefigure}{S\arabic{figure}}
\renewcommand{\theequation}{S\arabic{equation}}

\setcounter{figure}{0}
\setcounter{equation}{0}

\part{Supplemental Material}

\subsection{Carrier density profile of the simulated device}

\begin{figure}[b]
\subfigure[]{
\includegraphics[height=3.6cm]{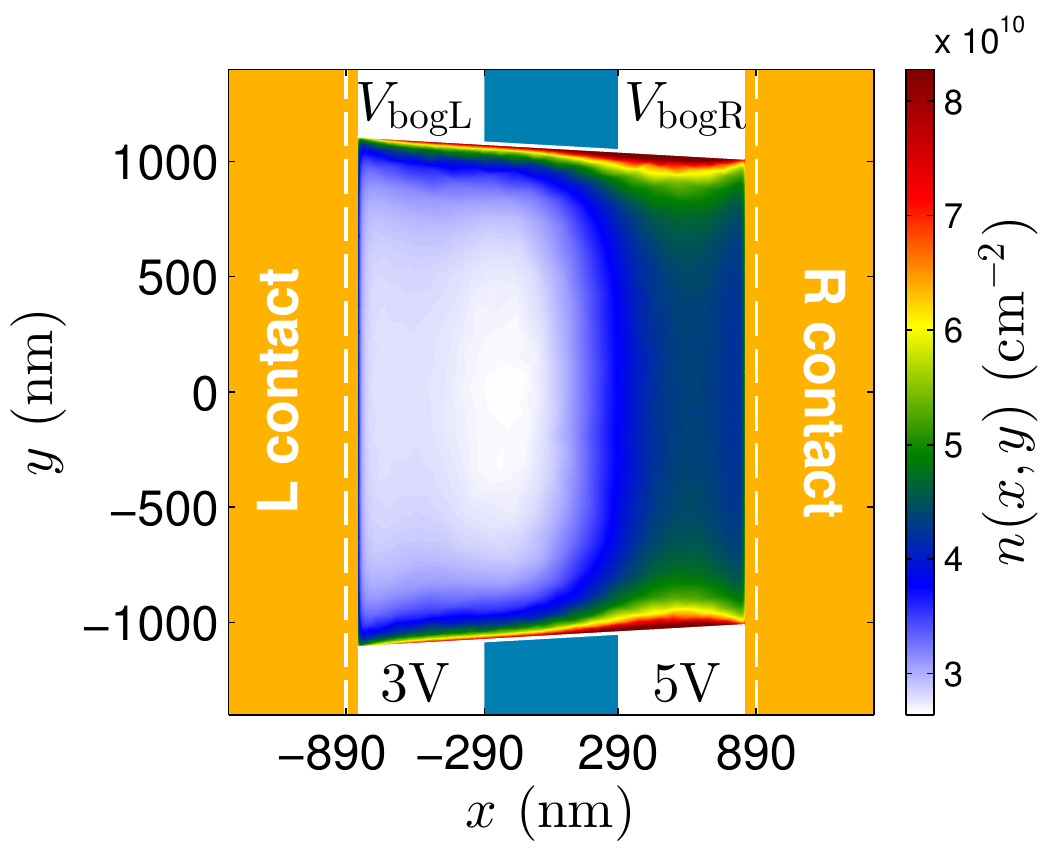}}
\subfigure[]{
\includegraphics[height=3.6cm]{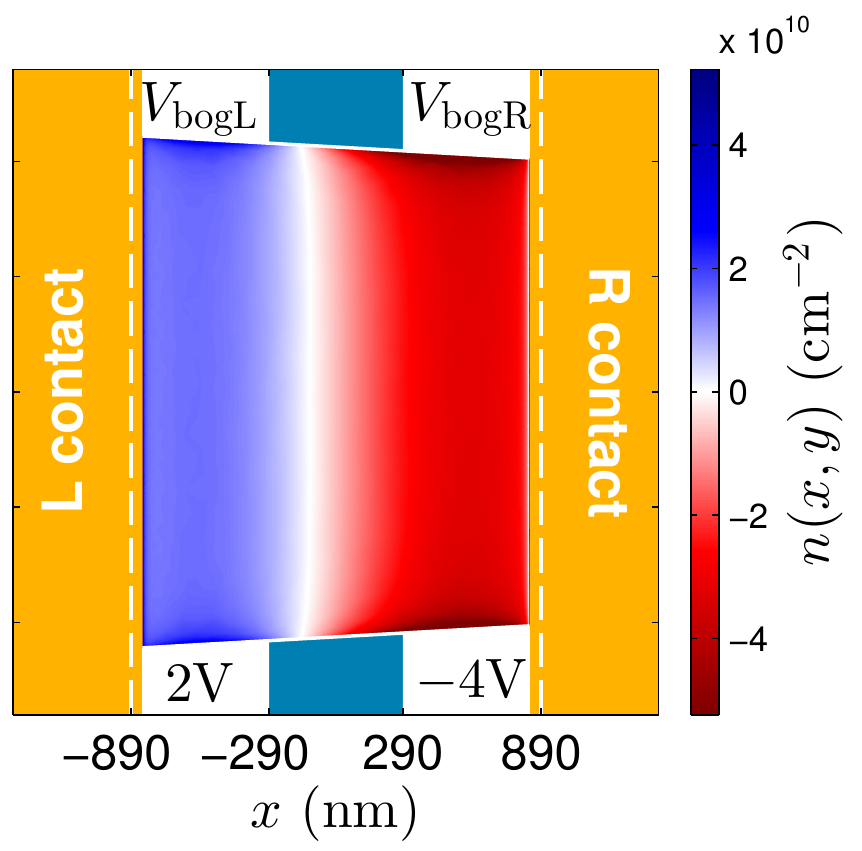}}
\caption{Examples of carrier density profiles $n(x,y)$ with (a) unipolar and (b) bipolar gate voltage configurations. Bottom gate voltages are indicated in respective plots. The geometry follows the design values of the experiment, and the shape of the graphene flake is estimated from an optical image of the real device. The width of the bottom gates is $600\unit{nm}$, and the white dashed lines indicate the edges of the bottom gates underneath the contacts.}
\label{figS1}
\end{figure}

As mentioned in the main text, the finite-element simulator FEniCS \cite{FEniCS} together with the mesh generator GMSH \cite{gmsh} are adopted to compute the self-partial capacitances \cite{Liu2013} of the individual metal contacts and bottom gates, $C_{\text{cL}},C_{\text{cR}},C_{\text{bogL}},$ and $C_{\text{bogR}}$, which are functions of two-dimensional coordinates $(x,y)$. The classical contribution to the total carrier density $n(x,y)$ is given by the linear combination $\sum_{i=\text{cL,cR,bogL,bogR}}(C_{i}/e)V_{i}$, where $V_{\text{bogL}}$ and $V_{\text{bogR}}$ are the left and right bottom gate voltages, respectively, and $V_{\text{cL}}$ and $V_{\text{cR}}$ are responsible for contact doping mainly arising from the charge transfer between the metal contacts and the graphene sheet. Since the experimental conditions are very similar to our previous work \cite{Rickhaus2013}, we adopt the same empirical value of $0.04\unit{V}$ for both $V_{\text{cL}}$ and $V_{\text{cR}}$. Total carrier density follows Ref.\ \onlinecite{Liu2013}. Two examples showing $n(x,y)$ profiles are given in Fig.\ \ref{figS1}, where zero intrinsic doping is assumed.

\subsection{Carrier-energy relation in the presence of magnetic field}

To compute the carrier density as a function of energy $E$ and magnetic field $B_{z}$ using the Green's function method, we consider an ideal graphene ribbon extending infinitely along the $\pm x$ axis. The retarded Green's function gives the total density of states of the supercell, $D(E,B_{z})=-(1/\pi )\func{Im}\func{Tr}G^{r}(E,B_{z})$, where we have explicitly denoted the dependence of the magnetic field $B_{z}$, which enters from the tight-binding Hamiltonian of the supercell. The carrier density in the zero temperature limit is given by integrating over the energy, $n(E,B_{z})=(2/A)\int_{0}^{E}D(E^{\prime },B_{z})dE^{\prime }$, where the factor $2$ accounts for the spin degeneracy and $A=N(3\sqrt{3}a^{2}/4)$ is the area of the supercell with $N$ the number of lattice sites within the supercell and $a=s_{f}a_{0}$ the lattice spacing. 

\begin{figure}[b]
\subfigure[]{
\includegraphics[height=4.8cm]{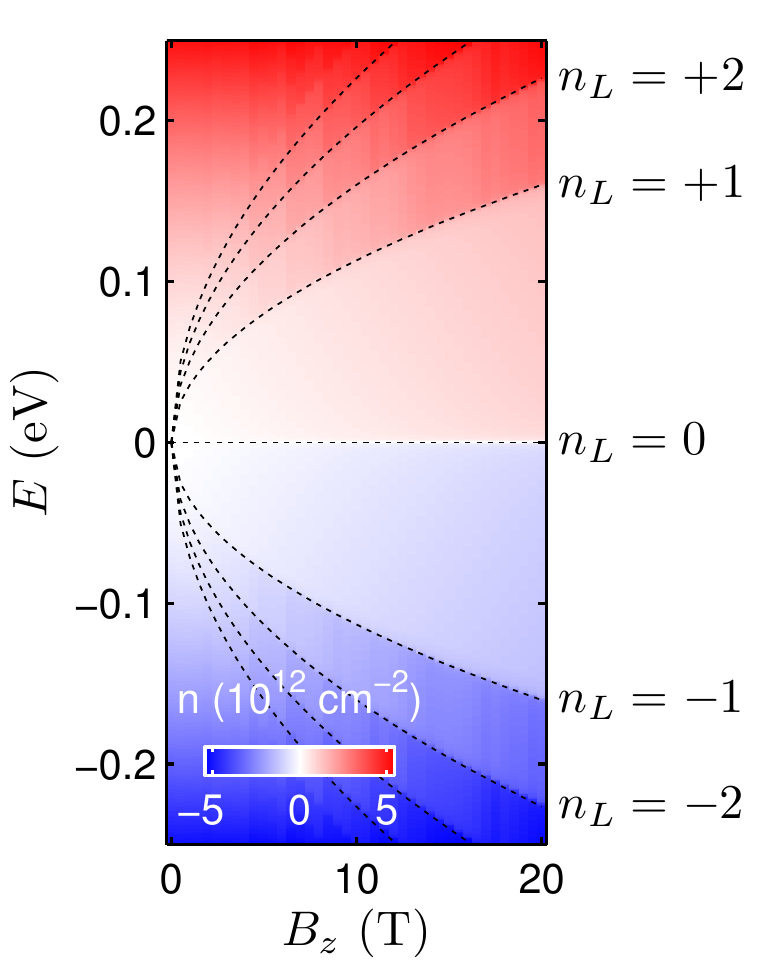}}
\subfigure[]{
\includegraphics[height=4.8cm]{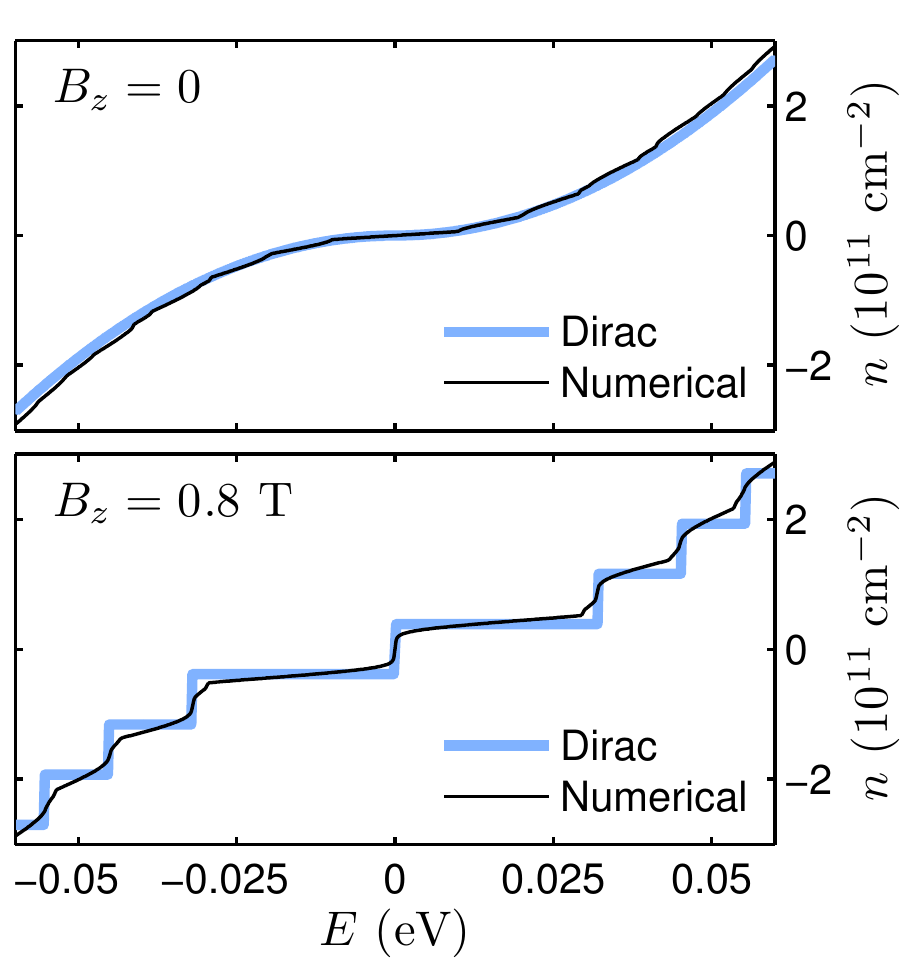}}
\caption{(a) Carrier density as a function of energy $E$ and magnetic field $B_{z}$, using an $s_{f}=4,N_{a}=101$ artificial armchair graphene ribbon (about 100 nm wide). The quantized carrier density is well described by the Landau level spectrum (dashed lines) given by Eq.\ \eqref{EnL}. (b) Carrier-energy relation at $B_{z}=0$ (upper panel) and $B_{z}=0.8\unit{T}$ (lower panel), using an $s_{f}=16,N_{a}=50$ ribbon (about 200 nm wide). The numerical results are compared with the Dirac model, Eq.\ \eqref{nE_Dirac} in the upper panel and Eq.\ \eqref{nEB_Dirac} in the lower panel.}
\label{figS2}
\end{figure}

An example for $n(E,B_{z})$ using a scaled armchair graphene ribbon with $s_{f}=4$ and $N_{a}=101$ (about 100 nm wide) is given in Fig.\ \ref{figS2}(a). With the increasing $B_{z}$, the emergence of the relativistic Landau level spectrum is clearly seen, which is well described by
\begin{equation}  \label{EnL}
\begin{split}
E_{n_{L}}&=\func{sgn}(n_{L})E_{1}\sqrt{|n_{L}|} \\
E_{1}&=\sqrt{2eB_{z}\hbar v_{F}^{2}}
\end{split}
,\quad n_L=0,\pm 1,\pm 2, \cdots.
\end{equation}%
Thus properly scaled graphene also correctly captures the half integer quantum Hall physics of real graphene.

In Fig.\ \ref{figS2}(b) we use another ribbon with $s_{f}=16$ and $N_{a}=50$ (about 200 nm wide) to compare the carrier-energy relation with and without magnetic field. For the $B_{z}=0$ case [upper panel in Fig.\ \ref{figS2}(b)], despite the ribbon nature of the considered artificial graphene, the $n(E)$ relation is basically consistent with the Dirac model,%
\begin{equation}
n_{\text{Dirac}}(E)=\func{sgn}(E)\frac{1}{\pi }\left( \frac{E}{\hbar v_{F}}%
\right) ^{2}.  \label{nE_Dirac}
\end{equation}%
For the $B_{z}=0.8\unit{T}$ case [lower panel in Fig.\ \ref{figS2}(b)], the numerical result exhibits quantized plateaus due to the emerging Landau levels. The plateaus are, however, not perfectly flat due to the level broadening of the density of states, which stems from the finite width of the considered ribbon, instead of temperature.

In the case of ideal infinite graphene, the density of states can be written as $D_{\text{Dirac}}(E,B_{z})=(4eB_{z}/h)\sum_{n_{L}}\delta (E-E_{n_{L}})$, where the prefactor accounts for the states each Landau level can accommodate and $E_{n_{L}}$ is given in Eq.\ \eqref{EnL}. Integrating $D_{\text{Dirac}}(E,B_{z})$ with respect to energy, one obtains a perfectly quantized carrier-energy relation%
\begin{equation}
n_{\text{Dirac}}(E,B_{z})=\frac{4eB_{z}}{h}\left( \func{sgn}(E)\left\lfloor
\frac{E^{2}}{E_{1}^{2}}\right\rfloor +\frac{1}{2}\right) ,
\label{nEB_Dirac}
\end{equation}%
where $\left\lfloor x\right\rfloor $ stands for the largest integer not greater than $x$ (known as the floor function in computer science) and $E_{1} $ is given in Eq.\ \eqref{EnL}. Compared to the numerical $n(E,B_{z})$ [lower panel in Fig.\ \ref{figS2}(b)], the ideal $n_{\text{Dirac}}(E,B_{z})$ given by Eq.\ \eqref{nEB_Dirac} is not suitable for describing the carrier-energy relation in finite-width graphene systems. Nevertheless, the formula confirms the correct trend of the numerical carrier-energy relation in the presence of magnetic field.

From the numerical $n(E)$ curve at a given $B_{z}$, such as that given in Fig.\ \ref{figS2}(b), the position of the highest filled energy level for a given carrier density, $E(n)$, is obtained, and the negative of it is the desired energy band offset for transport calculation, which is adopted in the simulations for Fig.\ 3(e) of the main text as well as the following unipolar quantum Hall regime.

\subsection{Unipolar quantum Hall data}

Figure \ref{figS3}(a) shows the measured unipolar conductance map $G(B_{z},V_{\text{bog}})$ with the two bottom gates connected together. By subtracting the contact resistance $R_{c}\approx 1080\unit{\Omega}$, the quantized conductance at low field up to $0.2\unit{T}$ is compared with the computed transmission function using $s_{f}=50$ scaled graphene in Fig.\ \ref{figS3}(b), in the presence of Anderson-type disorder with strength $U_{\rm dis}=3\unit{meV}$ (see the main text). Note that the color range in Fig.\ \ref{figS3}(b) is adjusted to highlight the conductance plateaus up to filling factor $\nu =\pm 14$. 

Despite the rather consistent Landau fan diagrams in both experiment and theory maps of Fig.\ \ref{figS3}(b), a closer look shows that the minimal $B_{z}$ required to quantize the conductance in the experiment is larger than that in the simulation possibly because of thermal fluctuations not considered in the calculations. In addition, the slopes of the fan-shaped plateaus indicate a slightly different charging efficiency between the experiment and the simulation, which we analyze in the following.

\subsection{Gate efficiency from the Landau fan diagram}

\begin{figure}
\subfigure[\label{fig4a}]{\includegraphics[height=4.5cm]{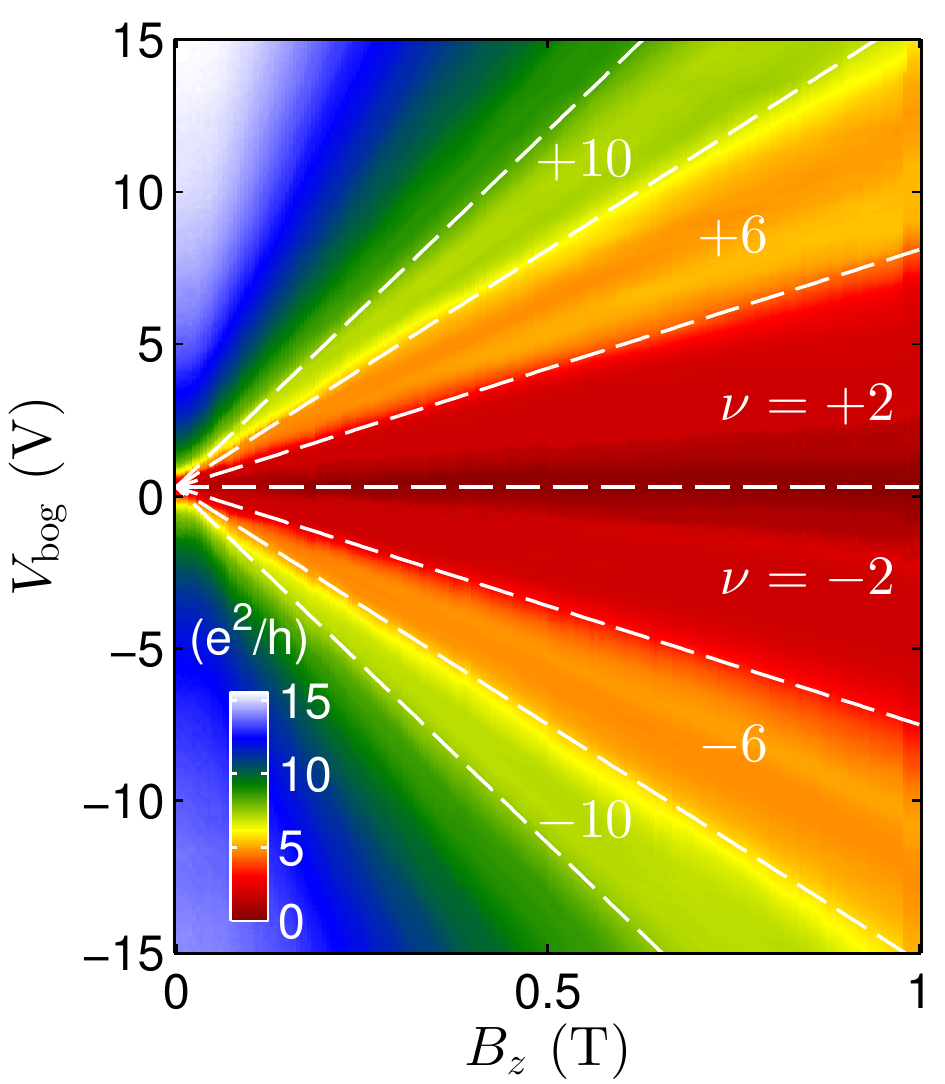}} %
\subfigure[\label{fig4b}]{\includegraphics[height=4.5cm]{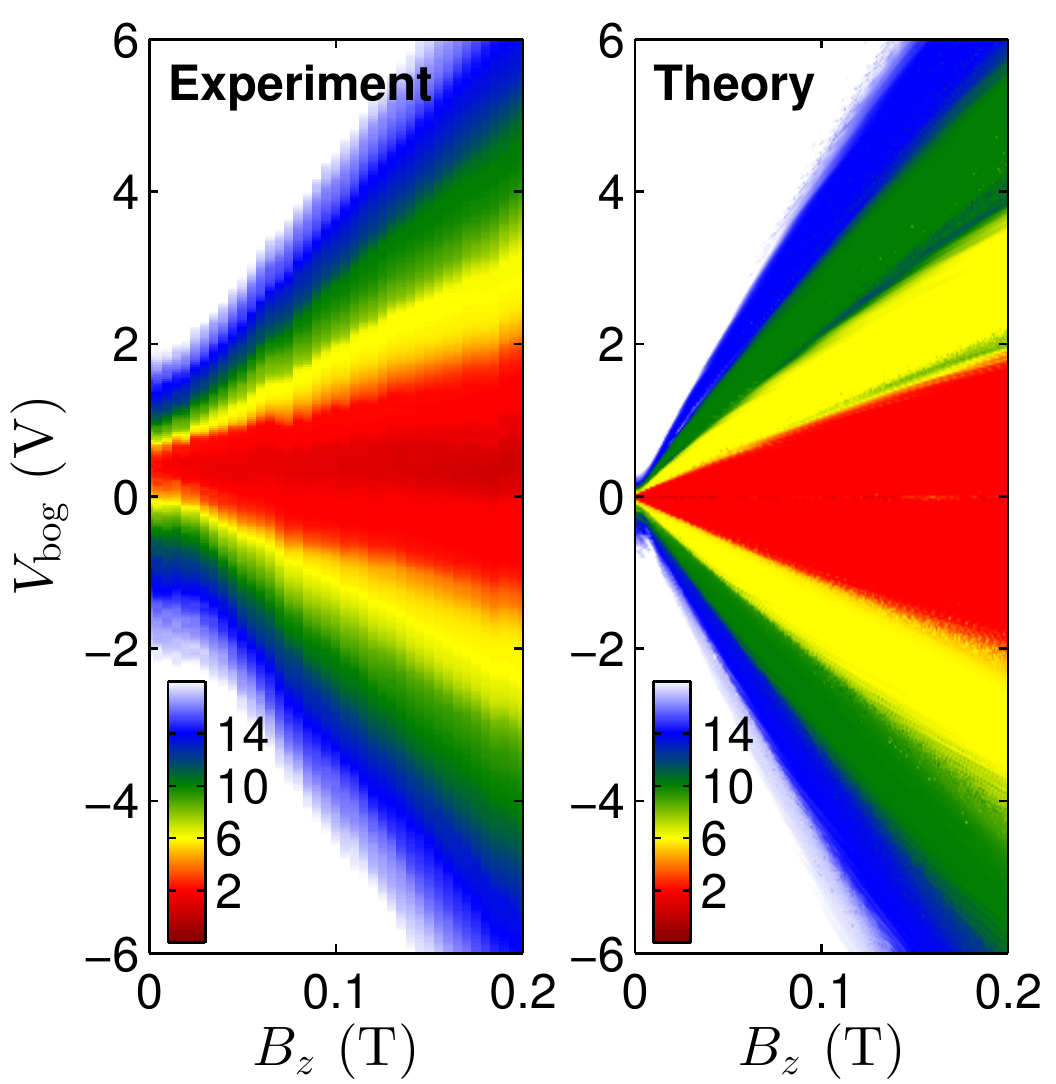}}
\caption{(a) Experimental data of the conductance measured with the two bottom gates connected to each other ($V_{\mathrm{bogL}}=V_{\mathrm{bogR}}=V_{\mathrm{bog}}$) and magnetic field $B_{z}$ sweep up to $1\unit{T}$. The lowest 6 quantized conductance plateaus labeled by filling factors $\protect\nu =\pm 2,\pm 6,\pm 10$ are separated by the fitting fan lines. (b) By subtracting the deduced contact resistance $R_{c}\approx 1080\unit{\Omega}$, the experimental data at low field is compared with the theory data of the computed transmission function $T$.}
\label{figS3}
\end{figure}

The pronounced quantized conductance plateaus reported in Fig.\ \ref{figS3}(a) allows for a precise evaluation of the gate efficiency. Let the average gate capacitance of the connected bottom gates be $\bar{C}_{g}$ and assume a uniform chemical doping of concentration $n_{0}$. Relating the mean carrier density given by $\bar{n}=n_{0}+\bar{C}_{g}V_{\text{bog}}$ and filling factor $\nu =\bar{n}/(eB_{z}/h)$ one finds 
\[V_{\text{bog}}=\frac{e\nu}{\bar{C}_{g}h}B_{z}-\frac{n_{0}}{\bar{C}_{g}}\equiv c_1\nu B_z+c_2.\] 
Thus on the measured field-gate map shown in Fig.\ \ref{figS3}(a), the slope of each fan line that separates two adjacent conductance plateaus $\nu -2$ and $\nu +2$ gives $c_{1}\nu =e\nu /\bar{C}_{g}h$ while the intersect at $B_{z}=0$ gives $c_{2}=-n_{0}/\bar{C}_{g}$. By fitting the experimental data at $\nu =0,\pm 4,\cdots $, we find $c_{1}=1.95\unit{V}\unit{T}^{-1}$ and $c_{2}=0.3\unit{V}$, which yield a gate efficiency 
\[\bar{C}_{g}=\frac{e}{c_{1}h}=1.24\times 10^{10}\unit{cm}^{-2}\unit{V}^{-1}\]
and a weak chemical doping 
\[n_{0}=-c_{2}\bar{C}_{g}=-3.72\times 10^{9}\unit{cm}^{-2},\] respectively.

\subsection{Comments on speed-up and bilayer graphene}

The strongly reduced memory demand brought by the scaling allows one to deal with previously prohibited micron-scale two-dimensional graphene systems. Even for computable systems, the speed-up can be seen in, e.g., the computation time $\Delta t$ for the lead self-energy that typically grows with the cube of the number of lattice sites within the lead supercell, i.e., $\Delta t\rightarrow \Delta t/s_{f}^{3}$ after scaling. Taking the illustrated 2.2-micron-wide graphene for example, $\Delta t$ is found to be $\sim 2.4\unit{s}$ on a single Intel Core i7 CPU for the artificial graphene scaled by $s_{f}=100$. For $s_{f}=1$, the time required to compute just a single shot of the self-energy, if the memory allowed, would be $\sim 2.4\times (100)^{3}\unit{s}$, which is almost a month.

The scaling also applies to bilayer graphene, as clearly seen from its energy spectrum given by \cite{McCann2013}
\begin{equation}\label{Ek BLG}
E(k)=\pm\sqrt{\frac{\gamma_{1}^{2}}{2}+\frac{U^{2}}{4}
+\hbar^{2}v_{F}^{2}k^2 \pm
\sqrt{\frac{\gamma_{1}^{4}}{4}+\hbar^2 v_{F}^{2}k^2
\left(\gamma_1^2+U^2\right)}},
\end{equation}
with $\gamma _{1}\approx 0.39\unit{eV}$ the interlayer nearest neighbor hopping and $U$ the asymmetry parameter responsible for the gap. The appearance of the product $ta$ in the dispersion \eqref{Ek BLG} after substituting $\hbar v_F=3ta/2$ clearly suggests that the scaling condition [Eq.\ (1) of the main text] also applies to bilayer graphene with $\gamma _{1}$ and $U$ left unaltered. Similar to the long wavelength limit [Eq.\ (2) of the main text] but due to the massive Dirac nature, the validity range of $s_{f}$ is more limited than the single-layer case. In the case of gapless bilayer graphene, we have $s_{f}\ll 6\pi t_{0}/[(2|E_{\max}|+\gamma _{1})^{2}-\gamma _{1}^{2}]^{1/2}$, which suggests $s_{f}\ll 50$ for the single-band transport ($|E_{\max }|\leq \gamma _{1}$). In the presence of magnetic field, the restriction of Eq.\ (3) in the main text remains true.

\bigskip\noindent
{\small Published version: \href{http://journals.aps.org/prl/abstract/10.1103/PhysRevLett.114.036601}{Phys.\ Rev.\ Lett.\ \textbf{114}, {036601} (2015)}}
\end{document}